\documentstyle[twoside,fleqn,espcrc2,psfig,subfigure]{article}
\title{Searching for $B_c$ mesons in the ATLAS experiment
at LHC\thanks{Work partially
supported by CICYT under contract AEN 93-0234 and by IVEI.
}}

\author{M. Gald\'on$^a$, R. P\'erez Ochoa$^{b,c}$ $\ $,
M. A. Sanchis-Lozano$^{a,b}$\thanks{Talk presented by M.A. Sanchis-Lozano in
the Conference Production and Decay of Hyperon, Charm and Beauty Hadrons.
Strasbourg, Sept. 1995. Email: mas@evalvx.ific.uv.es} and J.A. Valls$^{b,c}$\\
\vspace{0.5cm}
(a) Departamento de F\'{\i}sica Te\'orica \\
\vspace{0.1cm}
(b) Instituto de F\'{\i}sica Corpuscular (IFIC)\\
\vspace{0.1cm}
(c) Departamento de F\'{\i}sica At\'omica, Molecular y Nuclear \\
\vspace{0.1cm}
$\ $$\ $$\ $$\ $Dr. Moliner 50, E-46100 Burjassot, Valencia (Spain)}

\begin{document}

\begin{abstract}
\vspace{-6.5cm}
\large{
\begin{flushright}
  IFIC/95-62\\
  \today
\end{flushright} }
\vspace{6.cm}
We discuss the feasibility of the observation of the signal
from $B_c$ mesons in the ATLAS experiment at the LHC. In particular, we
address the decay mode $B_c{\rightarrow}J/\psi \pi$ followed by the leptonic
decay $J/\psi{\rightarrow}\mu^+\mu^-$, which should permit an accurate
measurement of the $B_c$ mass. We performed a Monte Carlo study of the signal
and background concluding that a precision of $\approx$ 1 MeV for the $B_c$
mass could be achieved after one year of running at $\lq\lq$low" luminosity.
The semileptonic decay $B_c{\rightarrow}J/\psi\ {\mu}^+{\nu}_{\mu}$ is also
considered for a possible extraction of ${\mid}V_{cb}{\mid}$.
\end{abstract}

\maketitle

\section{Introduction}
There is a general consensus in the scientific community \cite{aach} that
the scope of a future high-luminosity, high-energy hadron collider
like LHC should not be restricted to the hunting of the standard model
Higgs and its extensions, or the search for supersymmetry. Other topics
requiring lower luminosities like top and beauty physics deserve in their
own right a close \vspace{0.1in} attention.
\par

In particular, a lot of work has been recently devoted to the observation
of the $B_c$ meson regarding both the hadroproduction \cite{cheung} and
decay \cite{albiol} \cite{masl} (see references therein). Specifically, we
shall focus on the feasibility of its detection in the ATLAS experiment at
the LHC through some decay \vspace{0.1in} modes.

\section{$B_c$ signal}
At the center-of-mass energy $\sqrt{s}=14$ TeV, the cross-section for
beauty production is assumed to be $500$ $\mu$b leading to
$5{\times}10^{12}$ $b\overline{b}$ pairs per year-run ($10^7$ s) at a
luminosity of $\cal{L}\ {\approx}\ 10^{33}$ cm$^{-2}$ s$^{-1}$,
corresponding to an integrated luminosity of ${\sim}\ 10$ fb$^{-1}$. The
number of bottom pairs reduces, however, to $2.3{\times}10^{10}$
by requiring events with a triggering muon coming from either a $b$ or a
$\overline{b}$ under the kinematic cuts $p_{\bot}>6$ GeV/c and
${\mid}{\eta}{\mid}<2.2$ \vspace{0.1in} \cite{tp}.
\par
On the other hand, assuming that the $b$-quark fragmentation yields a $B_c$
or a $B_c^{\ast}$ with probability of the order of $10^{-3}$ \cite{cheung}, the
yield of $B_c$ mesons (not yet triggered) per year of running would be
roughly ${\simeq}\ 10^{10}$.
\vspace{0.1cm}
\subsection*{$B_c\ {\rightarrow}\ J/\psi\ \pi$ channel}
\vspace{0.2cm}
This exclusive channel followed by the leptonic decay of the
$J/\psi$ resonance into a pair of oppositely charged muons offers
several important advantages. First of all, it allows for the mass
reconstruction of the $B_c$ meson. Observe also that anyone of the two muons
can trigger the decay. Besides, it is very clean topologically with a common
secondary vertex for all three charged particles, two of them (the muons) with
the additional constraint of their invariant mass compatible with a
$J/\psi$. Furthermore, the expected branching fraction is not too small, about
$0.2\%$ \cite{lusi}, which combined with the branching
fraction of the leptonic decay of the resonance
$BR(J/\psi\ {\rightarrow}\ {\mu}^+{\mu}^-)\ {\simeq}\ 6\%$ \cite{pdg},
yields an overall branching fraction for the signal of $10^{-4}$. Thus, the
number of such events turns out to be ${\simeq}\ 10^6$ per year of
\vspace{0.1cm} running.
\par
A study has been performed in order to estimate the signal
detection efficiency and background for the
$B_c\ {\rightarrow}\ J/\psi({\rightarrow}\ {\mu}^+{\mu}^-)\ \pi$ channel. The
Monte Carlo employed for the signal simulation corresponds to a sample of
$B_c\ {\rightarrow}\ J/\psi\pi$ events while for the background we
have used a sample of inclusive $b$ muon decays generated in all the cases
with PYTHIA 5.7 \vspace{0.16cm}.
\par
Two types of \vspace{0.16cm}background were considered:
\footnote{Muons coming from semileptonic decays of long-lived
particles such as pions or kaons contribute in a negligible amount to
trigger rates. On the other hand $B$ decays into $J/\psi$ and a charged
particle would give an invariant mass quite below the $B_c$ mass so they
are of no concern}
\begin{itemize}
\item[a)] Combinatorial background due to muons from semileptonic decays of
$b\overline{b}$ pairs produced at the main interaction. Cascade
contributions such as $b{\rightarrow}c{\rightarrow}\mu$ are included as
well for random combinations with any other muon in the same event.
\item[b)] Contamination from prompt $J/\psi$'s
in combination with another charged hadron (interpreted as a pion)
from the main vertex. (In fact data released by Tevatron on the $J/\psi$
yield point out a production rate quite larger than initially expected
\cite{greco}.) Incorrect tracking may give rise to the
reconstruction of a (fake) secondary vertex, becoming a potential
source of a large amount of \vspace{0.12cm} background.
\end{itemize}
\par
In a first step, we imposed the following cuts on events based on
kinematic constraints:
\begin{itemize}
\item $p_{{\bot}min}(trig.\mu)=6$ GeV/c \, ; \\
$\mid\,{\eta}_{max}(trig.\mu)\mid =2.2$
\item $p_{{\bot}min}(\mu)=3$ GeV/c \, ; \, $\mid\,{\eta}_{max}(\mu)\,\mid =
2.5$
\item $p_{{\bot}min}(\pi)=1$ GeV/c \, ; \, $\mid\,{\eta}_{max}(\pi)\,\mid =
2.5$
\item $M_{{\mu}^+{\mu}^-}=M_{J/\psi}\ {\pm}\ 50$ MeV
\end{itemize}
\vspace{0.16cm}
The two first cuts correspond to the requirement of the 1st-level $B$ physics
trigger leading in our case to an efficiency of $\sim 15\%$ in triggering
one of the two muons from the $J/\psi$. We next take into account the
detection efficiency
for the signal after applying the rest of $p_{\bot}$ and $\eta$ cuts which
turns
out to be $\sim 21\%$. Setting the efficiency for muon identification as
$80\%$ and the track reconstruction as $95\%$ \cite{tp} we get a
combined detection efficiency of $\sim 2\%$ leading to an observable signal of
about $20,000$ events per \vspace{0.16cm} year of running.
\par
The last of the cuts described above constrains the two muons invariant
mass to be compatible (within two standard deviations \cite{tp})
with the nominal $J/\psi$ mass, thus drastically
reducing random combinations. However, background of class $b)$ can
potentially pass all the kinematic cuts by a large amount, so another type
of rejection is \vspace{0.1in} required.
\par
To this end, we adapted to our needs the vertex reconstruction (i.e.
vertex finding and fitting) routines of the LEP experiment DELPHI at
CERN \cite{bil}. The vertex fitting algorithm
provides as output the coordinates of the secondary vertex, the track momenta
re-evaluated with the vertex constraint and the goodness of the fit
by means of the total ${\chi}^2$ as well as the contribution of each
single track to it. In particular, we employed for background rejection
the three spatial coordinates and the ${\chi}^2$ for each fitted secondary
vertex formed by the two muons and the charged hadron (assumed to be a pion)
satisfying the above kinematic constraints. The distance
between the reconstructed vertex and the primary ($pp$) interaction point
was thereby determined. We shall refer to it as the decay
length even for background events of \vspace{0.1in} class b).
\par
Hence, candidate (either signal or background) events were
required to pass the following extra \vspace{0.16cm} cuts:
\begin{itemize}
\item total ${\chi}^2\ <\ {\chi}_0^2$
\item ${\chi}_i^2\ <\  \frac{{\chi}_0^2}{3}$ for each single track-$i$
\item decay length larger than $L_0$
\end{itemize}
\vspace{0.2cm}
where ${\chi}_0^2$, $L_0$ have to be optimized to remove the background
as much as possible but with a good acceptance for the signal. In our
analysis we found ${\chi}_0^2=8$ and \vspace{0.1in} $L_0=350$ $\mu$m.
\par
Figure 1 shows the reconstructed $(\mu^+\mu^-)_{J/\psi}\ \pi$
mass distribution for the expected signal above the surviving background
once all the cuts have been applied, for an
integrated luminosity of $10$ fb$^{-1}$ \cite{albiol}.
\par

In summary, we have found that the self-triggering weak decay
$B_c{\rightarrow}J/\psi\ \pi$, followed by the leptonic decay of the $J/\psi$
into two muons, could be clearly observed in the ATLAS detector at LHC. Under
rather conservative assumptions, a total number of ${\approx}\ 10,000$
signal events could be fully reconstructed after one year run,
corresponding to $10$ fb$^{-1}$ at $\lq\lq$low" luminosity
(${\approx}\ 10^{33}$ cm$^{-2}$ s$^{-1}$). This represents a signal to
background ratio of about $0.5$  with a statistical significance of
${\approx}\ 20$ standard deviations above a nearby almost flat
background. The foreseen mass resolution (standard deviation)
of the $B_c$ meson is about $40$ MeV representing a precision of about
1 MeV for the mass measurement.

\begin{figure}
\centerline{\psfig
{figure=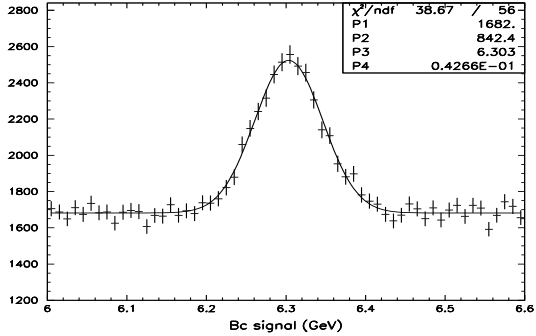,height=4.5cm,width=7.cm}}
\caption[\bf Figure 1]{Reconstructed $({\mu}^+{\mu}^-)_{J/\psi}\ \pi$ mass
distribution after cuts. The nominal value for the $B_c$ mass was
set equal to 6.3 GeV. The solid line corresponds to a linear+Gaussian fit}
\end{figure}

\subsection*{$B_c\ {\rightarrow}\ J/\psi\ \mu^+ \ {\nu}_{\mu}$ channel}
\vspace{0.2cm}
In spite of the fact that this channel does not permit
the measurement of the $B_c$ mass, its signature would be quite clean
experimentally when the $J/\psi$ decays into a pair of muons, providing
a three muon vertex. In the following we shall argue that in LHC experiments
this decay mode could provide a reliable extraction of
the mixing matrix element \vspace{0.1in} ${\mid}V_{cb}{\mid}$.
\par
The expected semileptonic branching ratio ${\simeq}\ 2\%$
\cite{lusi} together with the
leptonic branching ratio for the $J/\psi$ into two muons
${\simeq}\ 6\%$ \cite{pdg}, yields an overall branching fraction of
order $10^{-3}$. We next combine the last $BF$ with the acceptance of the muon
trigger ${\simeq}\ 15\%$ (where the three possible triggering muons per
decay have been taken into account.) Besides, the detection of the two
remaining particles (i.e. the non-triggering muons satisfying some $p_{\bot}$
and ${\eta}$ cuts \cite{nota}) amounts to an acceptance for signal events of
${\simeq}\ 12\%$. Finally, we assume an identification efficiency for each
muon of $0.8$ \cite{tp} yielding a combined value of \vspace{0.1in}
$0.8^3\ {\simeq}\ 0.5$.\par
Thereby, the expected number of useful $B_c$ decays \footnote{Semi-electronic
$B_c{\rightarrow}J/\psi e^+ \nu$ decays, followed by the $J/\psi$ decay
into ${\mu}^+{\mu}^-$ can be considered as well. In addition, if
the muon from the semi-muonic decay (1) gives the trigger, the
event can be selected with the mode $J/{\psi}{\rightarrow}\ e^+e^-$. The
trigger acceptance for the former is ${\simeq}\ 12\%$ whereas for the latter
is ${\simeq}\ 3\%$ \cite{nota}}
could reach several $10^5$ per year at low \vspace{0.1in} luminosity.\par
Such large statistics together with the foreseen precision in the muon momentum
measurement \cite{tp} should permit the experimental access to the kinematic
region near zero-recoil of charmonium (with respect to the $B_c$ meson) with
a good energy/momentum resolution. Moreover, very stringent cuts can be put
on events and hence, expectedly, background could be almost entirely removed
from the event sample permitting a very precise determination of the
$B_c$ lifetime. We explore all these issues in the \vspace{0.1in}
following section.

\subsubsection*{Theory inputs and determination of ${\mid}V_{cb}{\mid}$}
\vspace{0.2cm}
Next we examine in detail the possibility of extracting ${\mid}V_{cb}{\mid}$
from experimental data. Let us start by recalling \vspace{0.1in} that
\begin{eqnarray}
\frac{d{\Gamma}(B_c{\rightarrow}J/\psi\ {\mu}^+\nu)}{dw}\
=\nonumber \\
\frac{1}{\tau}\ \frac{dBr(B_c{\rightarrow}J/\psi\ {\mu}^+\nu)}{dw}
\end{eqnarray}
\vspace{0.06in}
\par
The differential branching ratio $dBr/dw$ can be written
\vspace{0.06in} as
\begin{eqnarray}
\frac{dBr(B_c{\rightarrow}J/\psi\ {\mu}^+\nu)}{dw}\ =\nonumber \\
\frac{1}{N(B_c)}\frac{dN(B_c{\rightarrow}J/\psi\ {\mu}^+\nu)}{dw}
\end{eqnarray}
where $dN/dw$ represents the number of semileptonic decays per unit
of $w$ for a certain integrated luminosity, once corrected by detection
efficiency; $N(B_c)$ is the corresponding total yield
of $B_c$'s produced for such integrated luminosity. (Notice that $\tau$ could
be determined accurately from the same collected sample of semileptonic
$B_c$ \vspace{0.1in} decays.)
\par
In order to get $N(B_c)$ it will be convenient to compare the
$B_c$ production rate to the prompt ${\psi}'$ yield in $pp$ collisions,
for the same integrated luminosity. \vspace{0.06in} Therefore
\begin{eqnarray}
N(B_c)\ =\ \frac{\sigma(pp{\rightarrow}B_c+X)}
{\sigma(pp{\rightarrow}{\psi}'+X)}\ \nonumber \\
{\times}\ \frac{N({\psi}'{\rightarrow}{\mu}^+{\mu}^-)}
{{\epsilon}_{{\psi}'}{\cdot}BF({\psi}'{\rightarrow}{\mu}^+{\mu}^-)}
\end{eqnarray}
where $N({\psi}'{\rightarrow}\mu^+\mu^-)$ stands for the number of prompt
${\psi}'$ to be experimentally detected through the muonic decay mode and
${\epsilon}_{{\psi}'}$ denotes its detection efficiency. Contamination
from weak decays of $B$ mesons into ${\psi}'$ states should be efficiently
removed by means of the vertex capability of the inner
\vspace{0.1in} detector.
\par
Therefore, we suggest normalizing the $B_c$ sample with the aid of the
${\psi}'$ yield through expression (3). The ${\psi}'$ state is probably
preferable to
the $J/\psi$ mainly because the former expectedly should not be fed down by
higher charmonium states \footnote{See Ref. \cite{close1} for an alternative
explanation of the observed ${\psi}'$ surplus found in Tevatron. However this
mechanism would require an uncomfortable large radiative $BF$ for higher
resonances}. Instead, data released by Tevatron shows that
the $J/\psi$ indirect production through ${\chi}_c$ intermediate states may
be also important. Let us also
remark that the production of the ${\psi}'$ resonance in $p\overline{p}$
collisions is more than one order of magnitude larger than initially
expected \cite{ma} \vspace{0.1in} \cite{greco}.
\par
At sufficiently large $p_{\bot}$ \footnote{We find an average
$p_{\bot}\ {\approx}\ 20$ GeV/c for $B_c$ mesons simply passing the
kinematics cuts on $p_{\bot}$ and ${\mid}{\eta}{\mid}$ of the decay muons
\cite{nota}}, one reasonably expects that the main
contribution to heavy quarkonia production comes from the splitting of
a heavy quark or a gluon. Indeed, even though the fragmentation process
is of higher order in ${\alpha}_s$ than the $\lq\lq$conventional" leading
order diagrams \cite{shuler}, the former is enhanced by powers of
$p_{\bot}/m_Q$ relative to the latter \cite{cheung} \cite{roy}. Although the
situation is still controversial in the literature, there are strong
indications
from very detailed calculations (see \cite{leike} \cite{greco2} and references
therein) that fragmentation indeed dominates for large enough $p_{\bot}$. In
the following, we first examine some color-singlet mechanisms, expected to
contribute largely to the high-$p_{\bot}$ inclusive production of prompt heavy
\vspace{0.1in} quarkonia.
\par
On the other hand, according to the set of papers in \cite{cheung},
perturbative QCD can provide a reliable calculation for the fragmentation
function of a high-$p_{\bot}$ parton into $B_c$ states with only few input
parameters:
the QCD running constant ${\alpha}_s$, the charm and bottom masses, and the
square of the radial wave function at the origin ${\mid}R(0){\mid}^2$.
Note that the largest uncertainty for normalization purposes comes from the
(third power of the) charm mass in the fragmentation function (see
\cite{cheung} for explicit expressions). However, we are only interested in
the \vspace{0.1in} ratio
\begin{equation}
r\ =\ \frac{\sigma(pp{\rightarrow}B_c+X)}{\sigma(pp{\rightarrow}{\psi}'+X)}
\end{equation}
where each cross section can be written as a convolution of the parton
distribution functions of the colliding protons, the cross section for the
hard subprocess leading to the fragmenting gluon or heavy quark and the
respective fragmentation function \cite{cheung}. Thus notice
that the fragmentation
functions $D_{\overline{b}{\rightarrow}B_c}$, $D_{g{\rightarrow}B_c}$ and
$D_{c{\rightarrow}{{\psi}'}}$, $D_{g{\rightarrow}{{\psi}'}}$ \cite{gluon},
which themselves are independent of the parton-level subprocesses, appear
combined as a ratio. Therefore, those uncertainties coming from the
common overall factor $m_c^3$ automatically
cancel each other. Moreover, those uncertainties introduced by
the parton distribution functions also should significantly diminish
in $r$ as well. Still one must evaluate the ratio
\begin{equation}
\frac{{\mid}R_{B_c}(0){\mid}^2}{{\mid}R_{{\psi}'}(0){\mid}^2}
\end{equation}
In fact there is the possibility of expressing the above ratio (aside trivial
factors) according to the general factorization analysis of \cite{bodwin2}
in a more rigorous way \vspace{0.1in} as
\begin{equation}
{\kappa}_0\ =\ \frac{<0{\mid}O_1^{B_c}(^1S_0){\mid}0>}
{<0{\mid}O_1^{{\psi}'}(^3S_1){\mid}0>}
\end{equation}
where $O_n^X$ are local four fermion operators\footnote{This implies the
application of NRQCD to the evaluation of the non-perturvative part of the
fragmentation function of a $b$ quark into a $b\overline{c}$ bound state.
Therefore four quark operators involve both $b$- and $c$-quark fields in this
case\cite{mannel}.}
and the matrix elements can be evaluated from NRQCD. A similar expression
holds for the $B_c^{\ast}$. (Moreover, the denominator can be determined from
the measured leptonic width of the \vspace{0.1in} ${\psi}'$.)\par
Recently, an additional color-octet fragmentation mechanism \cite{fleming}
has been suggested in order to reconcile the experimental results on inclusive
${\psi}'$ production with theoretical predictions. This mechanism assumes
the creation from gluon fragmentation of a $c\overline{c}$ pair in a
color-octet state, in analogy to ${\chi}_c$ production \cite{ma}. Then a new
nonperturbative parameter $H'_{8({\psi}')}$ is required, appearing
in $r$ combined as the dimensionless \vspace{0.1in} factor
\begin{equation}
{\kappa}_{8({\psi}')}\ =\ \frac{<0{\mid}O_8^{{\psi}'}(^3S_1){\mid}0>}
{<0{\mid}O_1^{{\psi}'}(^3S_1){\mid}0>}\ {\approx}\
\frac{2{\pi}m_c^2H'_{8({\psi}')}}
{{\mid}R_{{\psi}'}(0){\mid}^2}
\end{equation}
where we remark that it can be computed in terms of NRQCD matrix
\vspace{0.1in} elements \cite{bodwin2} \footnote{$H'_{8({\psi}')}$ can be
phenomenologically estimated from $B$ or $\Upsilon$ decays as well \cite{ma}
\cite{tot} \cite{bodwin}. On the other hand, according to a nonrelativistic
quark model $H'_8$ is related to a fictitious color-octet wave
function at the origin}.
\par
Furthermore, if we do not neglect the contribution to the total
$B_c$ production of those orbital excitations like $P$-wave states, new
fragmentation functions $D_{\overline{b}{\rightarrow}\overline{b}c(P)}$
come into play \cite{cheung2}. Accordingly, some new ${\kappa}_i$ factors
will appear which can be conveniently expressed, for instance in
analogy to Eqs. (6) and (7), as:
\begin{eqnarray}
{\kappa}_{1(B_c)}\ =\ \frac{1}{\hat{m}^2}\
\frac{<0{\mid}O_1^{B_c}(P_J){\mid}0>}{<0{\mid}O_1^{B_c}(^1S_0){\mid}0>}
\nonumber \\
{\propto}\ \frac{8{\pi}\hat{m}^2H_{1(\overline{b}c)}}
{{\mid}R_{B_c}(0){\mid}^2}
\end{eqnarray}
and
\begin{eqnarray}
{\kappa}_{8(B_c)}\ =\ \frac{1}{\hat{m}^2}\
\frac{<0{\mid}O_8^{B_c}(^3S_1){\mid}0>}{<0{\mid}O_1^{B_c}(^1S_0){\mid}0>}
\nonumber \\
{\propto}\ \frac{8{\pi}\hat{m}^2H'_{8(\overline{b}c)}}
{{\mid}R_{B_c}(0){\mid}^2}
\end{eqnarray}
where $\hat{m}$ is the reduced mass of the $b$ and $c$ quarks. The
long-distance $H_1$ parameter \cite{bodwin} is related to the square of
the derivative of the color-singlet wave function at the
\vspace{0.1in} origin.
\par
Higher $B_c$ resonances like $D$-wave states might be further taken
into account, introducing new nonperturbative parameters
involving higher derivatives of the wave function at the origin, or the
NRQCD matrix element \vspace{0.1in} analogues.
\par
In sum, the evaluation of such ${\kappa}_i$ parameters would allow one to
complete the computation of the ratio $r$ and thereby to obtain
from the experimental measurement of the ${\psi}'$ yield, the total number
of $B_c$ events for an integrated luminosity with the aid of
expression \vspace{0.1in} (3).
\par
On the other hand, the hadronic transition $B_c{\rightarrow}J/\psi$
involves two heavy-heavy systems whereby heavy quark spin symmetry should
be still valid as a first order
approach \cite{masl} \cite{mannel}. This permits the introduction
of the analogue of the Isgur-Wise function for transitions involving doubly
heavy hadrons, denoted by ${\eta}_{12}(v_1{\cdot}v_2)$. At the non-recoil point
$v_1=v_2$, we shall \vspace{0.06in} write
\begin{equation}
<\ J/\psi{\mid}\ A^{\mu}\ {\mid}B_c\ >\ =\
2\ {\eta}_{12}(1)\ \sqrt{m_1m_2}\ {\varepsilon}_2^{\ast\mu}
\end{equation}
where $A^{\mu}=\overline{c}{\gamma}^{\mu}{\gamma}_5b$ stands for
the axial-vector current and ${\varepsilon}_2^{\mu}$ represents the
four-vector polarization of the \vspace{0.1in} $J/\psi$.\par
Following similar steps as in the $B$ decay into $D^{\ast}$ \cite{neu2}, we
\vspace{0.06in} write
\begin{eqnarray}
{\lim}_{w{\rightarrow}1}\frac{1}{\sqrt{w^2-1}}\
\frac{d{\Gamma}(B_c{\rightarrow}{J/\psi}\ {\mu}^+{\nu})}{dw}\ =\nonumber \\
\frac{G_F^2}{4{\pi}^3}\ (m_1-m_2)^2\ m_2^3\ {\eta}_{12}^2(1)\
{\mid}V_{cb}{\mid}^2
\end{eqnarray}
\par
It is of key importance from the theoretical side in our proposal, the
existence of a single form factor in the above expression to be determined
theoretically in a rigorous manner.
Note that ${\eta}_{12}(1)$ may be interpreted in first approximation
as the overlap of the initial and final hadron wave functions \cite{masl}
\cite{mas}. Moreover, due to the intrinsic nonrelativistic nature of its heavy
constituents, nonrelativistic QCD on the lattice or even
refined potential based models of hadrons, could be applied
in order to get ${\eta}_{12}$ and its slope at zero recoil
in a reliable \vspace{0.1in} way.\par
This fact in conjunction with the presumed accuracy of the $J/\psi$
momentum measurement suggests determining ${\mid}V_{cb}{\mid}$ in a similar
way as in the method proposed by Neubert \cite{neu1} through the
\vspace{0.1in} $\overline{B}{\rightarrow} D^{\ast}\ell\overline{\nu}$ decay.
\par

\begin{figure}
\centerline{\hbox{
\psfig{figure=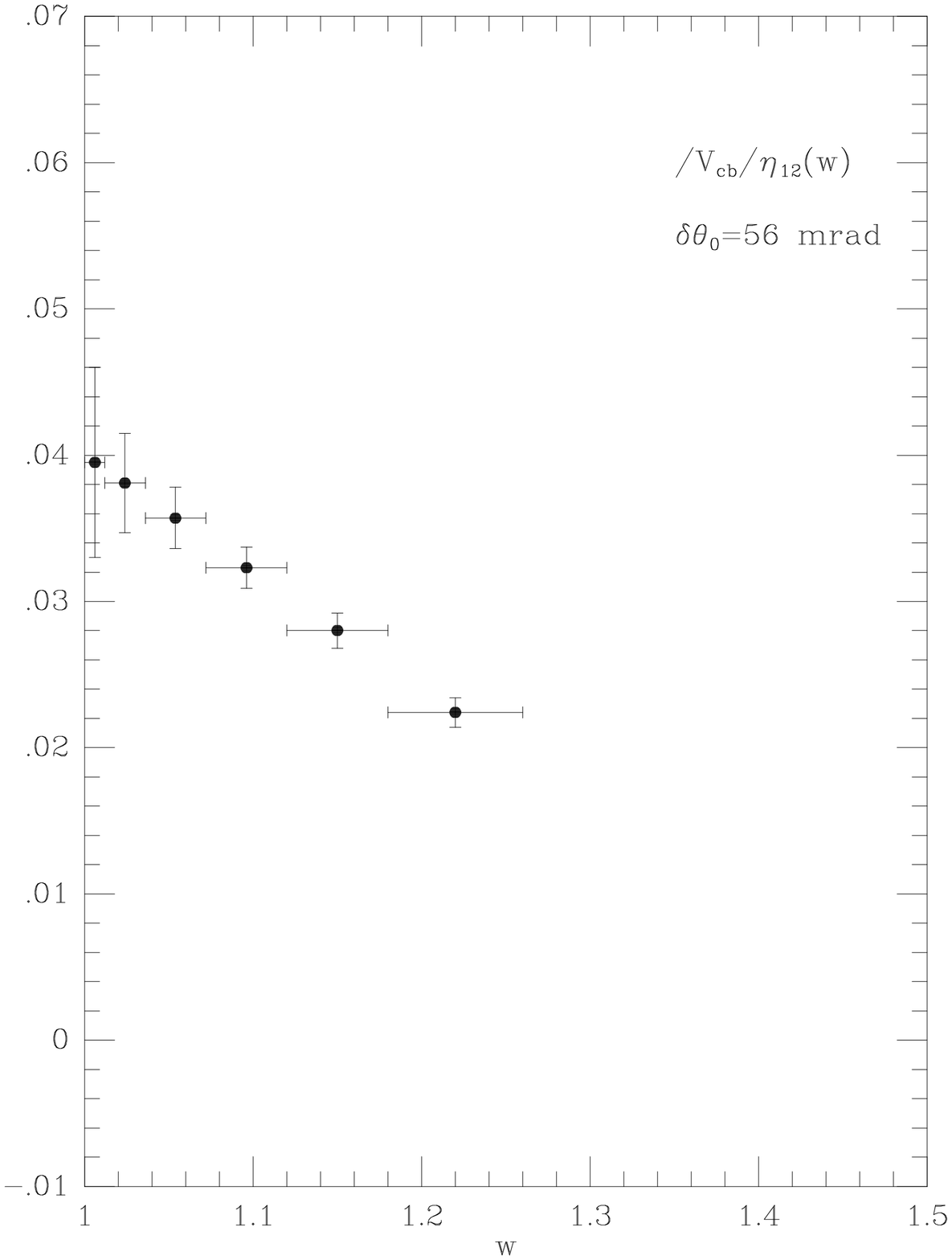,height=5.cm,width=6.cm}
}}
\centerline{\hbox{
\psfig{figure=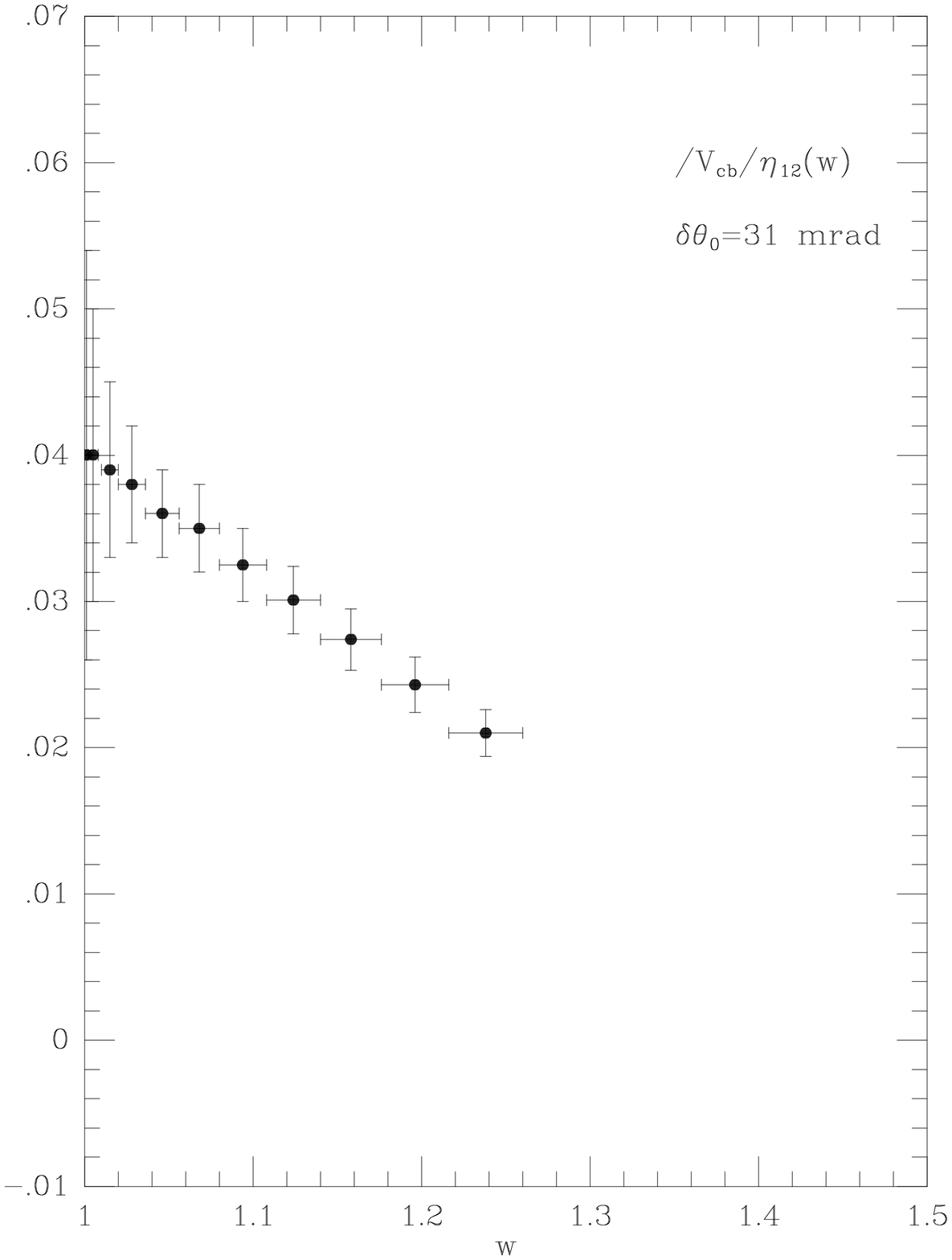,height=5.cm,width=6.cm}
}}
\caption[\bf Figure 2]{Hypothetical ${\eta}_{12}(w){\cdot}{\mid}V_{cb}{\mid}$
distributions of the decay $B_c{\rightarrow}J/\psi \mu^+\nu$ for a
$500$ events sample tentatively assigned to an integrated luminosity
of $20\ fb^{-1}$ (i.e. two years of running at low luminosity). It was assumed
an spatial resolution for the $B_c$ flight-path reconstruction of
${\sigma}\ {\approx}\ 100\ \mu$m and its lifetime is allowed to vary
in the interval ${\tau}=0.5-1$ ps. Horizontal
bars show systematic errors basically limiting the
number of experimental points. Vertical bars take into account the
statistical fluctuations of the sampled events in the bin (distributed
according to the expected differential rate for a pseudoscalar to vector
semileptonic decay \cite{neu2}). Points
line up with ${\eta}'_{12}$ set equal to $-2$. For more details see
Ref. \cite{maga}}
\end{figure}

For the purpose of illustration, we have depicted in figure 2
some expected typical points and error bars for the decay
$B_c{\rightarrow}J/\psi({\rightarrow}{\mu}^+{\mu}^-)\ {\mu}^+\nu$ for
a definite prediction of the ratio of production cross sections
$r$ in Eq.(4).
\par

\vspace{0.1cm}

Under this condition, let us remark that the uncertainty on
$w$ (i.e the horizontal error bars) is of systematic nature, mainly
coming from the $B_c$ flight-path reconstruction. With regard to the vertical
coordinate ${\eta}_{12}(w){\cdot}{\mid}V_{cb}{\mid}$, we have assumed that
its uncertainty is essentially statistical \vspace{0.1in}\cite{albiol}.
\par

\vspace{0.1cm}

It is important to stress that the quantity to be
experimentally measured with a large accuracy from
$B_c{\rightarrow}J/\psi\ {\mu}^+\nu$ decays can be written as the linear
combination:
\[ r^{1/2}\ {\times}\ {\eta}_{12}(1)\ {\times} {\mid}V_{cb}{\mid} \]
\par
In summary, we suggest to keep an open mind on the possibility of an
alternative determination of ${\mid}V_{cb}{\mid}$ from semileptonic decays
of $B_c$ mesons at LHC. Even if at the time LHC will start to run, B
factories would have already provided a (still) more precise value
of ${\mid}V_{cb}{\mid}$ than at present via the semileptonic
$B$ decay, an independent determination of it should bring a valuable
cross-check. On the other
hand, a lot of activity is being devoted to the analysis of inclusive
production of prompt charmonium resonances at Fermilab. Therefore we conclude
that if the fragmentation approach to describe the production of
heavy quarkonia at the large transverse momentum domain is accurately
tested (and tuned) from experimental data, $B_c$ semileptonic
decays could be competitive with the $B$ \vspace{0.1in} ones.\par
Alternatively, one can turn the question round and
consider ${\mid}V_{cb}{\mid}$ as a well-known parameter thus verifying
QCD calculations, either at the production or at the weak decay level.
Indeed, fragmentation into heavy quarkonia offers an interesting check
of perturbative QCD and a deep insight into the nonperturbative dynamics in
the hadronic formation. Besides, a precise knowledge of the $B_c$
production rate is a necessary condition for the experimental
measurement of the absolute branching fraction of any of its decay
modes.

\end{document}